\documentclass[aps,prd,twocolumn,superscriptaddress,floatfix,nofootinbib,
               amsmath,amssymb,reprint]{revtex4-2}

\usepackage{graphicx}
\usepackage{bm}
\usepackage{xcolor}
\usepackage{xspace}
\usepackage{lineno}
\usepackage[normalem]{ulem}   

\graphicspath{{figures/}}

\newcommand{\snn}{\sqrt{s_{\mathrm{NN}}}}
\newcommand{\AAcrit}{A_{\mathrm{th}}}
\newcommand{\dvdeta}{dv_{1}/d\eta}
\newcommand{\dSdy}{dS/dy}
\newcommand{\dNdeta}{dN_{\mathrm{ch}}/d\eta}
\newcommand{\Kn}{\mathrm{Kn}}   
\newcommand{\fp}{f_{p}}         
\newcommand{\Npart}{\langle N_{\mathrm{part}} \rangle}
\newcommand{\xiE}{\xi_{\mathrm{Entropy}}}
\newcommand{\xiP}{\xi_{\mathrm{Participant}}}
\newcommand{\xiG}{\xi_{\mathrm{Geometric}}}

\begin{document}
\title{Driver and damping of the directed-flow response in heavy-ion collisions}

\author{Kishora Nayak}
\affiliation{Department of Physics, Panchayat College, Bargarh 768028, Odisha, India}

\author{Vipul Bairathi}
\email[Corresponding author: ]{vipul.bairathi@gmail.com}
\affiliation{Instituto de Alta Investigaci\'on, Universidad de Tarapac\'a, Casilla 7D, Arica 1000000, Chile}

\date{\today}

\begin{abstract}
The initial-state geometry plays a crucial role in driving directed flow, while the dissipative response of the medium dampens it. Both of these factors influence how the directed-flow slope varies with system size. We developed a method to differentiate between these driving and damping effects on charged-hadron directed flow in O+O, Cu+Cu, Ru+Ru, Au+Au, and U+U collisions at $\sqrt{s_{\mathrm{NN}}} = 200$~GeV using an improved version of the string-melting AMPT model. We formulated three scaling observables based on the entropy density, the number of participants, and the mass number. A dimensionless ratio was constructed, revealing the threshold mass number $A \approx 35$ in central collisions from the entropy and participant scaling, indicating the onset of collective behavior. We constructed a kinetic-theory Knudsen-number ($\mathrm{Kn}$) map to analyze the contributions of the initial-state driver, which grows as $\Kn^{\kappa}$ with $\kappa \approx 2$, and a final-state viscous damping of the hydrodynamic response with characteristic scale $\Kn_0 \approx 0.27$ for the directed flow slope. This damping scale is found to be about a factor of 2.5 smaller than the $\Kn_0 \approx 0.7$ extracted from the elliptic flow. Furthermore, we determined the ratio of shear viscosity to entropy density, $\eta/s$, to be between 0.10 and 0.20, using an alternative method that does not rely on fitting flow harmonics.
\end{abstract}

\maketitle
\section{Introduction}
\label{sec:intro}
A hot, dense, and strongly interacting state of matter known as quark-gluon plasma (QGP) is produced in the collisions of atomic nuclei at experimental facilities such as the Relativistic Heavy Ion Collider (RHIC) and the Large Hadron Collider (LHC)~\cite{Arsene:2004fa,Back:2004je, Adams:2005dq, Adcox:2004mh}. The collective expansion of this matter is reflected in the azimuthal angle distribution of the produced hadrons relative to the collision symmetry plane~\cite{Heinz:2013th}. This distribution can be decomposed into Fourier harmonics as,
\begin{equation}
  \frac{dN}{d(\phi-\Psi_{n})} \;\propto\; 1 + 2\sum_{n=1}^{\infty} v_n
  \cos\!\big[n(\phi-\Psi_n)\big],
\end{equation}
where $\Psi_n$ is the angle of the $n^{\mathrm{th}}$-order symmetry plane and $v_n$ are the corresponding flow harmonics~\cite{Voloshin:1994mz, Poskanzer:1998yz}. The first harmonic $v_1$, directed flow, encodes both the tilt of the produced matter and the asymmetry between hard and soft particle production along the longitudinal direction~\cite{Bozek:2010aj, Adil:2005qn, Chatterjee:2017ahy}. Therefore, the pseudo-rapidity ($\eta$) dependence of directed flow, $v_1(\eta)$, and its slope at midrapidity, $dv_{1}/d\eta|_{\eta=0}$, is one of the most direct experimental measures to study early-time dynamics and the equation of state of the produced QGP matter~\cite{Brachmann:1999xt, Snellings:1999bt, Stocker:2004qu, Adamczyk:2017nxg, Adam:2019wnk, Abelev:2008jga}.

Azimuthal correlations relative to the collision symmetry plane, commonly referred to as collective flow, provide important information about the properties of the matter produced during high-energy heavy-ion collisions. The experimental observation of large collective flow in heavy-ion collisions, such as Au+Au and Pb+Pb, has been considered as a signature of the formation of the QGP. Hydrodynamic and transport models effectively describe the observed collective flow of particles at low transverse momentum, indicating nearly perfect fluid behavior of the QGP in these collisions, with the shear viscosity to entropy density ratio very close to the universal lower limit of $1/4\pi$ as predicted by AdS/CFT correspondence~\cite{Kovtun:2004de}. Recent results from small collision systems, in both symmetric (p+p and light A+A) and asymmetric (p/d/$^{3}\mathrm{He}$+A), at the LHC and RHIC have reported significant collective flow~\cite{AlICE:HighMult_flow_pp_pPb,ALICE:2019multipart,ALICE:2021XeXe,ALICE:OO_NeNe,ATLAS:2026OONeNe,Adare:2018toe}. The similarities in collective flow observed in both small and large collision systems suggest a common underlying physics mechanism; however, it remains a subject of active investigation. Research continues to examine factors, including the impact of initial-conditions, hydrodynamic behavior, transport dynamics, and hadronic re-scattering, to understand better the dependence of QGP formation on collision system size and processes such as thermalization.

A key question raised by the discovery of azimuthal correlations in small collision systems such as high multiplicity $p$+$p$, O+O, Ne+Ne, and Xe+Xe collisions~\cite{Shi:2008bs, ALICE:2021XeXe, GrosseOetringhaus:2024}, is the minimum system size at which the medium produced in heavy-ion collisions can be described as a thermalized, hydrodynamically expanding fluid. To address this question, a traditional geometric scaling of flow harmonics by nuclear mass number ($A^{1/3}$) has long been used to compare different collision systems~\cite{Schmidt:1993pi, Reisdorf:2012zz, Andronic:2003rx, Lang:1991eq, Chance:1997kg}. Recent transport model calculations have examined a wide range of collision systems, from O+O to U+U at $\snn = 200$ GeV. These studies have shown that directed flow and its slope, $\dvdeta$, do not scale with $A^{1/3}$ for both low- and high-$p_{T}$ charged hadrons~\cite{Nayak:2025epjc}. The deviation from simple geometric scaling suggests that the dynamics and properties of the matter may vary across different collision systems. Consequently, this scaling violation motivates the search for a threshold system size. Below this threshold, the matter produced in these collisions transitions from partonic to hadronic-dominated matter. More recently, a broad phenomenology of small-system collectivity has been comprehensively reviewed~\cite{GrosseOetringhaus:2024}, which emphasizes the absence of a single framework that describes the transition across all system sizes and does not attempt to quantify the size at which collective behavior sets in. 

In this work, we present a study of charged-hadron directed flow slope across five different collision systems (O+O, Cu+Cu, Ru+Ru, Au+Au, and U+U) at $\snn =$ 200 GeV using the string melting version of a multiphase transport (AMPT-SM) model with an improved quark coalescence mechanism. We tested, for the first time, a new quantitative approach to three independent directed-flow scaling observables based on the entropy density, the number of participants, and the mass number using a transport model. A dimensionless ratio is then constructed to extract the threshold mass number, marking the onset of collective behavior. More importantly, beyond the extraction of threshold mass number, the same system-size scan of directed flow slope can be recast in terms of the kinetic-theory Knudsen number ($\Kn$). Plotting the scaled directed flow against $\Kn$ can be used to distinguish between the hydrodynamic and free-streaming regimes of the medium produced in the relativistic heavy-ion collisions. The calculations are particularly compelling because they provide a data-driven approach to separate the initial-state \emph{driver} (due to the tilted, dipole-like fireball geometry) and final-state \emph{damping} (due to the viscous hydrodynamic response) components of the directed flow slope. The viscous corrections to the ideal hydrodynamics response of the medium in terms of shear viscosity to entropy density ratio can be obtained through $\Kn$. Therefore, the threshold system size required for hydrodynamic behavior to emerge would be provided by mapping the directed flow as a function of system size.

The paper is organized as follows. Section~\ref{sec:model} provides a brief overview of the AMPT-SM model, detailing its input parameters, as well as the process of generating and selecting collision events. Section~\ref{sec:framework} provides a comprehensive description of the methodology used in this study. This includes the three directed-flow scaling observables, the normalized ratio $\xi(A)$, the functional forms used to fit $\xi(A)$, and the threshold mass-number extraction. This section also presents the approach used to map the Knudsen number to the scaled directed-flow slope. In section~\ref{sec:results}, we present the findings related to directed flow scaling and provide estimates of the threshold mass number $\AAcrit$ based on the fit to $\xi(A)$, followed by the Knudsen-number map and the decomposition of the directed-flow response into its initial-state driver and viscous damping components. Lastly, Section~\ref{sec:summary} summarizes and discusses the findings of this work along with an outline for future outlook.

\section{Model description}
\label{sec:model}
\subsection{AMPT-SM Model}
\label{sec:model-ampt}
A multi-phase transport model, AMPT, is widely used in relativistic heavy-ion collisions~\cite{Lin:2004en}. It describes the full evolution of the collisions, from initial conditions to the final-state particles. The AMPT model provides a successful description of bulk and flow observables in heavy-ion collisions at both RHIC and  LHC~\cite{Lin:2004en,He:2017tla}. In this study, we employed the string-melting version of the AMPT model, incorporating a new quark coalescence approach introduced in Ref.~\cite{He:2017tla}. This improvement enhances the simultaneous description of light-flavor hadron yields and elliptic flow at top RHIC energy. The model evolution proceeds in four stages: (i) Generation of the initial partons from melted strings, produced by the HIJING event generator; (ii) Elastic rescatterings among the produced partons through Zhang's parton cascade  (ZPC) model~\cite{Zhang:1999bd}; (iii) Hadronization of the freeze-out partons through an improved quark coalescence mechanism as outlined in Ref.~\cite{He:2017tla}; and (iv) Hadronic rescattering, which occurs within a relativistic transport (ART) model. The input parameters of the model include a parton-parton cross-section $\sigma_{\mathrm{p}} = 1.5$~mb and a hadron cascade time $t_{\mathrm{max}} = 0.4$~fm/$c$, consistent with those used in Ref.~\cite{Nayak:2025epjc}.  

\subsection{Collision Systems and Nuclear Geometry}
\label{sec:model-systems}
We have simulated collisions of five different symmetric systems: $^{16}$O+$^{16}$O, $^{63}$Cu+$^{63}$Cu, $^{96}$Ru+$^{96}$Ru, $^{197}$Au+$^{197}$Au and $^{238}$U+$^{238}$U at $\snn = 200$~GeV using the AMPT-SM model. This selection of collision systems spans a wide range of nuclear mass number ($A$) and charge number ($Z$), enabling a comprehensive study of system-size dependence through scaling tests of the directed flow observable. The nuclear density profile of the colliding nuclei is simulated using a deformed Wood-Saxon distribution function within the model as,
\begin{equation}
  \rho(r,\theta) = \frac{\rho_0}{1 + \exp\left[\big\{r-R(\theta)\big\}/a\right]},
\label{eq:WS}
\end{equation}
with the angular dependent radius parameter, 
\begin{equation}
  R(\theta) \;=\; R_0\,
  \big[\,1 + \beta_2\,Y_{20}(\theta)\big].
\label{eq:WS_def}
\end{equation}
The values of nuclear radius $R_0$, surface diffuseness $a$, and deformation parameters $\beta_2$ used in this work are listed in Table~\ref{tab:WS}. These values are taken from our previous study of $v_1$ at the same beam energy~\cite{Nayak:2025epjc} and are consistent with the recent isobaric parameterizations~\cite{Sinha:2023plb, Zhao:2022prc, Giacalone:2021prl}. 
\begin{table}[!htbp]
\caption{Wood-Saxon and deformation parameters of the five symmetric collision systems~\cite{Nayak:2025epjc,Sinha:2023plb,Zhao:2022prc,Giacalone:2021prl}.}
\label{tab:WS}
\begin{ruledtabular}
\begin{tabular}{lccccc}
Nucleus    & $A$  & $Z$  & $R_0$ (fm) & $a$ (fm) & $\beta_2$\\
\hline
$^{16}$O   & 16   & 8    & 2.608      & 0.513    & 0         \\
$^{63}$Cu  & 63   & 29   & 4.214      & 0.586    & 0         \\
$^{96}$Ru  & 96   & 44   & 5.090      & 0.460    & 0.162     \\
$^{197}$Au & 197  & 79   & 6.380      & 0.535    & 0         \\
$^{238}$U  & 238  & 92   & 6.810      & 0.550    & 0.280     \\
\end{tabular}
\end{ruledtabular}
\end{table}

\subsection{Event Selection}
\label{sec:model-event}
The events generated using the AMPT-SM model for each collision system are categorized into various centrality intervals, ranging from 0\% to 80\% of the total cross-section. Centrality is defined on an event-by-event basis, using the multiplicity distribution of charged pions, kaons, protons, and antiprotons within $|\eta| < 0.5$, following the standard experimental procedure. This reference multiplicity distribution is divided into the different centrality classes. The mean number of participating nucleons $\Npart$ for each centrality interval is computed by counting the projectile and target nucleons that undergo at least one binary collision according to the Glauber Monte Carlo approach embedded within the AMPT model.

\section{Analysis Method}
\label{sec:framework}
\subsection{Directed Flow and Slope}
\label{sec:model-v1slope}
The rapidity-odd directed flow of charged hadrons produced in the final state is calculated relative to the reaction plane angle ($\Psi_{RP}$) using the following equation,
\begin{equation}
v_1(\eta) = \langle \cos(\phi - \Psi_{RP}) \rangle,
\label{eq:v1def}
\end{equation}
where $\phi$ is the azimuthal angle of the produced particles. The obtained charged hadron $v_1(\eta)$ is fitted with a cubic polynomial within $|\eta| < 1.2$~\cite{Nayak:2025epjc},
\begin{equation}
v_1(\eta) = F\eta + F'\eta^{3}.
\label{eq:v1cubic}
\end{equation}
The linear coefficient $F$ represents the mid-rapidity $v_1$-slope as,
\begin{equation}
F \equiv \frac{dv_1}{d\eta}\Big|_{\eta=0}.
\label{eq:dv1deta}
\end{equation}
In this analysis, charged hadrons within the $p_{T}$ range of $0.2 < p_T < 2.0$~GeV/$c$ are selected to suppress the contributions from high-$p_T$ particles and to isolate the bulk component of $v_1$ from the hard component.

\subsection{Scaling Techniques}
\label{sec:framework-norms}
This study examines three scaling techniques that utilize different normalization factors: (i) entropy density ($\dSdy$), (ii) mean number of participants ($\Npart$), and (iii) nuclear mass number ($A^{1/3}$). The entropy density of the produced matter is a natural candidate for scaling the bulk observable. In the hydrodynamic framework, the system's response is determined by energy density, pressure, entropy density, and temperature through the equation $\varepsilon + P = sT$~\cite{Heinz:2013th, Romatschke:2017ejr}. Moreover, the same volume can accommodate varying amounts of entropy depending on the collision energy and the number of participating nucleons~\cite{Heinz:2013th, Romatschke:2017ejr, Kovtun:2004de, Bjorken:1982qr, Kolb:2003dz}. The second choice of scaling by $\Npart$ is relevant when the flow signal is the linear superposition of an elementary per-source contribution. Lastly, the choice of scaling by nuclear mass number is based on the assumption that the relevant variable is related to the geometry of the colliding nuclei, which is appropriate if the flow signal builds up over a path length proportional to the nuclear radius.

We begin by formulating an observable based on the entropy scaling as follows,
\begin{equation}
S(A) \equiv \frac{F}{\dSdy},
\label{eq:SA}
\end{equation}
and we hypothesize that 
\begin{equation}
S(A) = \mathrm{constant},~\text{if}~A > \AAcrit,
\label{eq:RA}
\end{equation}
where $\AAcrit$ represents the threshold mass number below which thermalization remains incomplete, and the hydrodynamic response of the system has not been fully achieved. To test the hypothesis stated in Eq.~(\ref{eq:RA}), we compare it with two other scaling techniques that are commonly used in heavy-ion research: participant and geometric scaling. The observable based on participant scaling is defined as,
\begin{equation}
P(A) \equiv \frac{F}{\Npart},
\label{eq:PA}
\end{equation}
and it is sensitive to the number of participating nucleons from the initial state. The observable based on geometric scaling, 
\begin{equation}
G(A) \equiv \frac{F}{A^{1/3}},
\label{eq:GA}
\end{equation}
is sensitive to the size of the colliding nuclei. 

The three scaling scenarios will help us determine whether the system-size dependence of the $v_1$-slope is influenced by the final-state entropy density, the initial participant nucleons, or the geometrical length scale. If the corresponding scaling held across all system sizes, the above ratios would remain constant. However, if the scaling only holds above a certain threshold ($A > A_{th}$), then the ratios takes on a sigmoidal shape. The ratios decreases from the smallest system ($A = 16$) and eventually flattens out on a plateau at larger $A$ values. The plateau thus represents the regime where the scaling is restored, in accordance with the hypothesis of Eq.~(\ref{eq:RA}). The rise towards the smaller system reflects the breakdown of the scaling. In this case, the half-maximum point of the sigmoid function defines the threshold mass number $\AAcrit$.

Based on the three scaling approaches mentioned above, we construct a dimensionless ratio, defined as,
\begin{equation}
  \xi(A) \equiv \frac{R(A)}{R(\mathrm{O{+}O})},
  \qquad R\in\{S, P, G\},
  \label{eq:xiSC}
\end{equation}
where the smallest collision system, O+O ($A=16$), is taken as the reference. The $\xi(A)$, in Eq.~\ref{eq:xiSC} corresponding to the symbol $R = S$, $P$, and $G$, are called $\xiE$, $\xiP$, and $\xiG$, respectively. By construction $\xi(16) = 1$ for all three choices. The double-ratio nature of $\xi(A)$ allows for its experimental measurement, which effectively cancels out systematics associated with the detector.

\subsection{Entropy Proxy}
\label{sec:framework-entropy}
The initial state of relativistic nucleus-nucleus collisions can be characterized by several factors, such as energy density and temperature, with higher energy densities leading to increased entropy production and density. As particles collide and interact in a rapidly expanding system, the entropy density rises until thermal equilibrium is reached. The multiplicity of final-state particles serves as a key indicator of the entropy generated during these collisions, as higher multiplicity indicates a higher entropy density. 

In this study, we obtained rapidity entropy density, denoted as $\dSdy$, on an event-by-event basis, using the standard charged-hadron multiplicity $\dNdeta$ at $|y| <$ 1 as a proxy from the AMPT model, 
\begin{equation}
\dSdy = c \times \dNdeta.
 \label{eq:dSdy_proxy_def}
\end{equation}
The central value $c$ is set to 3.8 based on the thermal-model fits from Ref.~\cite{Kolb:2003dz} at top RHIC energy. The numerical value of $c$ is influenced by factors such as the chemical freeze-out temperature, the baryon chemical potential, and the contribution of mesonic and baryonic resonances. Standard thermal model implementations suggest a range of $c$ between 3.6 and 4.0 at $\snn = 200$~GeV. Note that the extracted value of $\AAcrit$ is independent of $c$, as $c$ cancels out exactly in the dimensionless ratio $\xi(A)$.

\subsection{S-curves and Power-law}
\label{sec:framework-fits}
In this subsection, we discuss the form of power-law function and a family of three S-curves. The power-law function is defined as follows, 
\begin{equation}
f^R(A) = \alpha\,A^{-\gamma},
\label{eq:powerlawF}
\end{equation}
where both the $\alpha$ and $\gamma$ are free parameters. The ratios $R(A)$ from Eqs.~\ref{eq:SA}-~\ref{eq:GA} are fitted with the power-law function to extract the parameter $\gamma$. The exponent $\gamma$ indicates the strength of the scaling violation; a larger $\gamma$ signifies that the per-source $v_1$ signal is suppressed more rapidly as the system size increases.

The three S-curves that parameterize $\xi(A)$ are represented by the Tanh function, the Error function, and the Gompertz function as follows,
\begin{equation}
\small f^{T}(A) = \xi_{\min} + \frac{(1-\xi_{\min})}{2} \left[1 - \tanh\left(\frac{A-\AAcrit}{w}\right)\right],
\label{eq:tanhfit}
\end{equation}

\begin{equation}
\small f^{E}(A) = \xi_{\min} + \frac{(1-\xi_{\min})}{2} \left[1 - \mathrm{erf}\left(\frac{A-\AAcrit}{w\sqrt{2}}\right)\right],
\label{eq:erffit}
\end{equation}

\begin{equation}
\small f^{G}(A) = \xi_{\min} + (1-\xi_{\min})\exp\left[-\exp\left(\frac{A-\AAcrit}{w}\right)\right],
\label{eq:gompertzfit}
\end{equation}
each parameterized by an asymptotic floor $\xi_{\min}$. The $\AAcrit$ is the threshold mass number at which $\xi$ crosses the half-maximum point, $(1 + \xi_{\min})/2$, and $w$ is the transition width. The $\xi_{\min}$ is a value in the limit of large A, above the threshold, which represents the plateau region as discussed in Sec.~\ref{sec:framework-norms}. The $w$ is a measure of the  width over which the transition from hadronic to hydrodynamic regime can be identified. A small value of $w$ corresponds to a sharp onset of collectivity, whereas a larger value of $w$ ($\gtrsim \AAcrit$) can hinder the transition region and indicate only a gradual crossover. This approach allows us to identify the threshold mass number above which the medium begins to exhibit characteristics of a strongly interacting hydrodynamic medium. 

\subsection{Knudsen-number map}
\label{sec:framework-knudsen}
The Knudsen number ($\Kn$) is a dimensionless quantity defined as the ratio of a microscopic length ($\lambda$) to a macroscopic scale ($L$). The $\lambda$ corresponds to the mean free path of particles, whereas $L$ represents the characteristic physical length scale of the system. It is inspired by kinetic theory and helps us understand the behavior of the medium created during relativistic nuclear collisions. The Knudsen number can be evaluated as~\cite{Drescher:2007cd,Lacey:2006bc,Nugara:2025}, 
\begin{equation}
\Kn = L\left[\frac{\fp\sigma_{p}}{\pi\tau_0}\left(\frac{dN_{ch}}{d\eta}\right)\right]^{-1},  
\label{eq:Kn}
\end{equation}
where $\fp$, $\sigma_p$, and $\tau_0$ are the parton multiplicity factor, the parton-parton cross-section, and the thermalization time, respectively. The charged particle multiplicity $\dNdeta$ is obtained from the AMPT model. The parameter $L$ in Eq.~\ref{eq:Kn} is the effective fireball radius, which is calculated as $L = r_0\,(0.5N_{\mathrm{part}})^{1/3}$ with $r_0 =$1.2~fm~\cite{Kolb:2003dz,Heinz:2013th}.

A very small Knudsen number ($Kn\ll1$) suggests that particle interactions occur frequently compared to their mean free path. In this case, the medium behaves like a fluid, which allows for hydrodynamic descriptions. Conversely, a large Knudsen number ($Kn\gg1$) characterizes a medium of relatively free particles without frequent collisions, where local thermal equilibrium may not be complete. The Knudsen number around unity ($Kn \approx 1$) identified as a transitional regime between hydrodynamic and free streaming. The Knudsen number can also be used to determine $\eta/s$ as it is directly proportional to it.

\section{Results}
\label{sec:results}
\subsection{Entropy Density}
\label{sec:dSdy}
\begin{figure}[!htbp]
\centering
\includegraphics[width=0.95\columnwidth,height=0.7705\columnwidth]{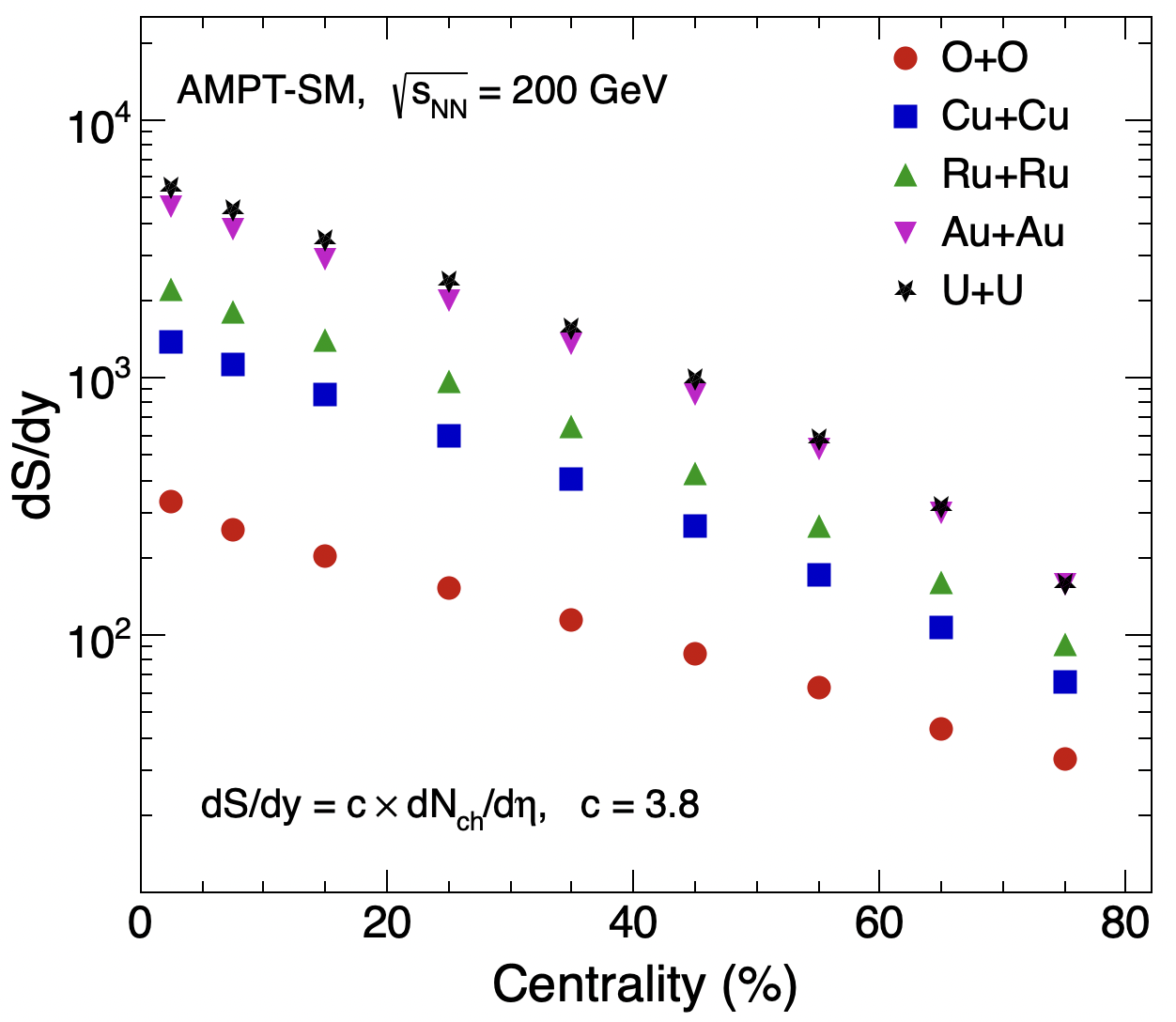}
\caption{Entropy density $\dSdy$ as a function of centrality (\%) for different collision systems at $\snn = 200$~GeV from the AMPT-SM model.}
\label{fig:dSdy}
\end{figure}

The entropy density $\dSdy$, derived using the charged hadron multiplicity at mid-rapidity according to Eq.~\ref{eq:dSdy_proxy_def} with $c = 3.8$ from the AMPT-SM model, is presented in Fig.~\ref{fig:dSdy}. The produced entropy follows the canonical mass ordering as: $(\dSdy)_{\mathrm{O+O}} \ll (\dSdy)_{\mathrm{Cu+Cu}} < (\dSdy)_{\mathrm{Ru+Ru}} < (\dSdy)_{\mathrm{Au+Au}} \lesssim (\dSdy)_{\mathrm{U+U}}$ within a given centrality interval. The largest collision system, U+U, generates approximately fifteen times more entropy density than O+O in the most central collisions. This significant variation in system dynamics provides a test for the entropy scaling hypothesis. Further, the entropy density within each system decreases monotonically by an order of magnitude from the most central to the peripheral collisions. The $dS/dy$ values shown in Fig.~\ref{fig:dSdy} are used to calculate entropy scaling of the directed flow slope. In the following section, we will present the results of the entropy scaling of $\dvdeta$ in comparison to other scaling approaches.

\subsection{Directed-Flow Scaling}
\label{sec:scaling-comparison}
\begin{figure*}[!htbp]
\centering
\includegraphics[width=0.92\textwidth]{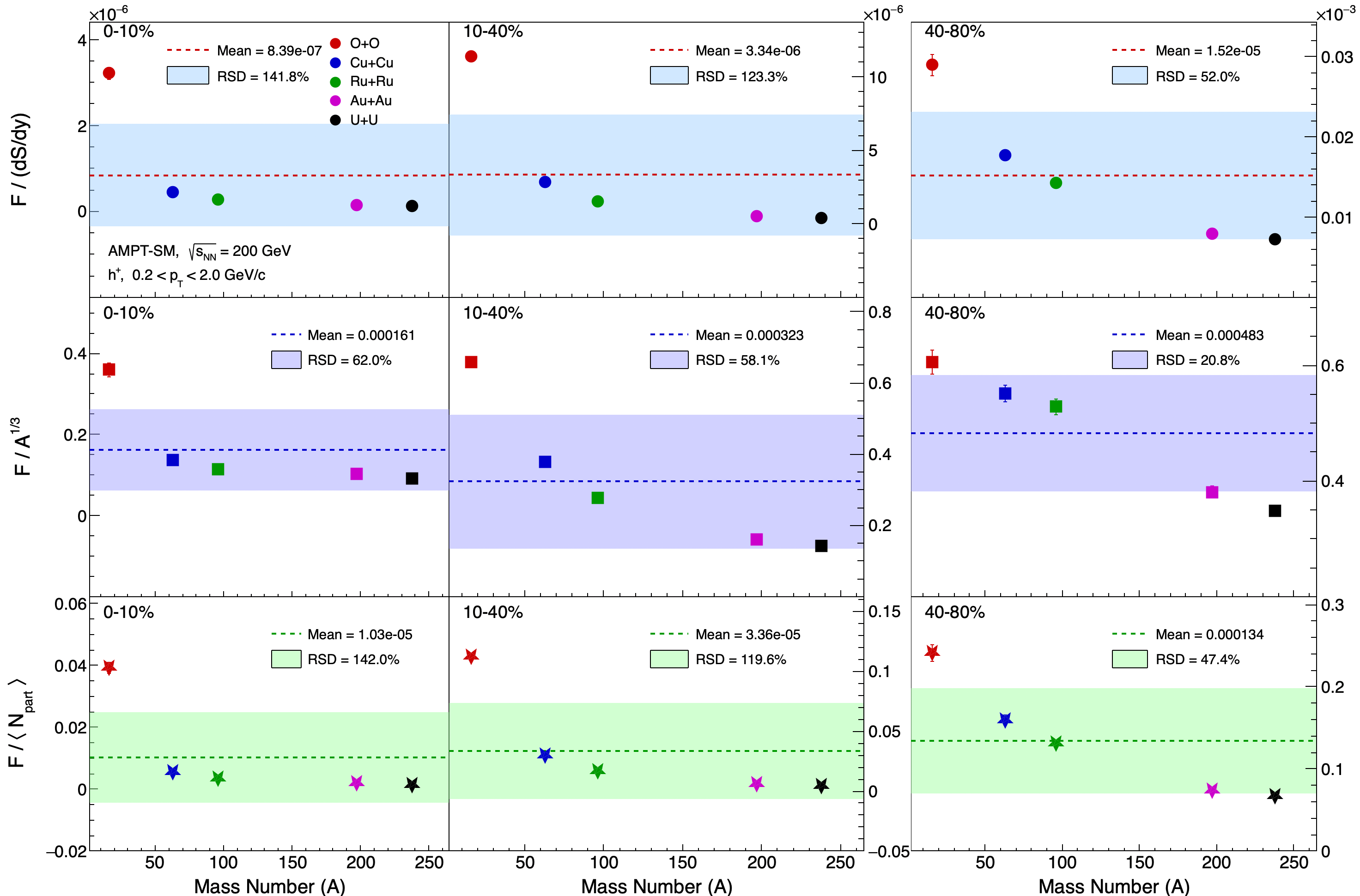}
\caption{The scaled $v_1$-slope ($F$) of charged hadrons at low $p_T$ ($0.2 < p_T < 2.0$~GeV/$c$) as a function of mass number $A$ for different centrality ranges: 0-10\%, 10-40\%, and 40-80\%. The top row shows entropy scaling $F/(\dSdy)$, the middle row shows geometric scaling $F/A^{1/3}$, and the bottom row shows participant scaling $F/\Npart$. The dashed horizontal lines show the global mean, while shaded bands show the relative standard deviation (RSD) across all colliding systems.}
\label{fig:RSD3x3}
\end{figure*}

Figure~\ref{fig:RSD3x3} shows mass number ($A$) dependence of the three scaling observables: $F/(\dSdy)$, $F/\Npart$, and $F/A^{1/3}$ for the charged hadron $v_1$-slope in central, mid-central, and peripheral collisions at $\snn =$ 200 GeV from the AMPT-SM model. We observed that all three scalings exhibit a similar qualitative trend; they decrease as the mass number increases. To quantify the observed trend, we use the relative standard deviation (RSD), which is calculated as RSD (\%) $= 100\times($SD$/$mean). In this equation, the mean represents the average value of the corresponding observable across all five collision systems and SD represents the standard deviation. The RSD shows a similar magnitude for entropy and participant scaling compared to geometric scaling. The RSD values for entropy-scaling observable changes from $142\%$ to $52\%$ from central to peripheral collisions, respectively. A similar centrality dependence is also observed for the other two scaling observables. 

In central collisions (0-10\%), for mass number A $\gtrsim$ 50, the scaled $v_1$-slope remains nearly constant and slightly below the global mean. However, for the lighter systems (A $<$ 50), we observe a significant deviation from the mean. The scaled slope values fall outside the RSD band, consistent with the hypothesis shown in Eq.~\ref{eq:RA}, which states that above the thermalization threshold, the flow observable in heavy-ion collisions follows near-common scaling, whereas below the threshold, it deviates from the scaling. This suggests that the collective behavior of the medium formed in lighter collision systems differs from that in heavier systems. In the peripheral collisions (40-80\%), the three scaled observables show a smooth transition from heavier to lighter colliding systems. 

\subsection{Suppression Exponent $\gamma$}
\begin{figure}[!htbp]
\centering
\includegraphics[width=0.95\columnwidth]{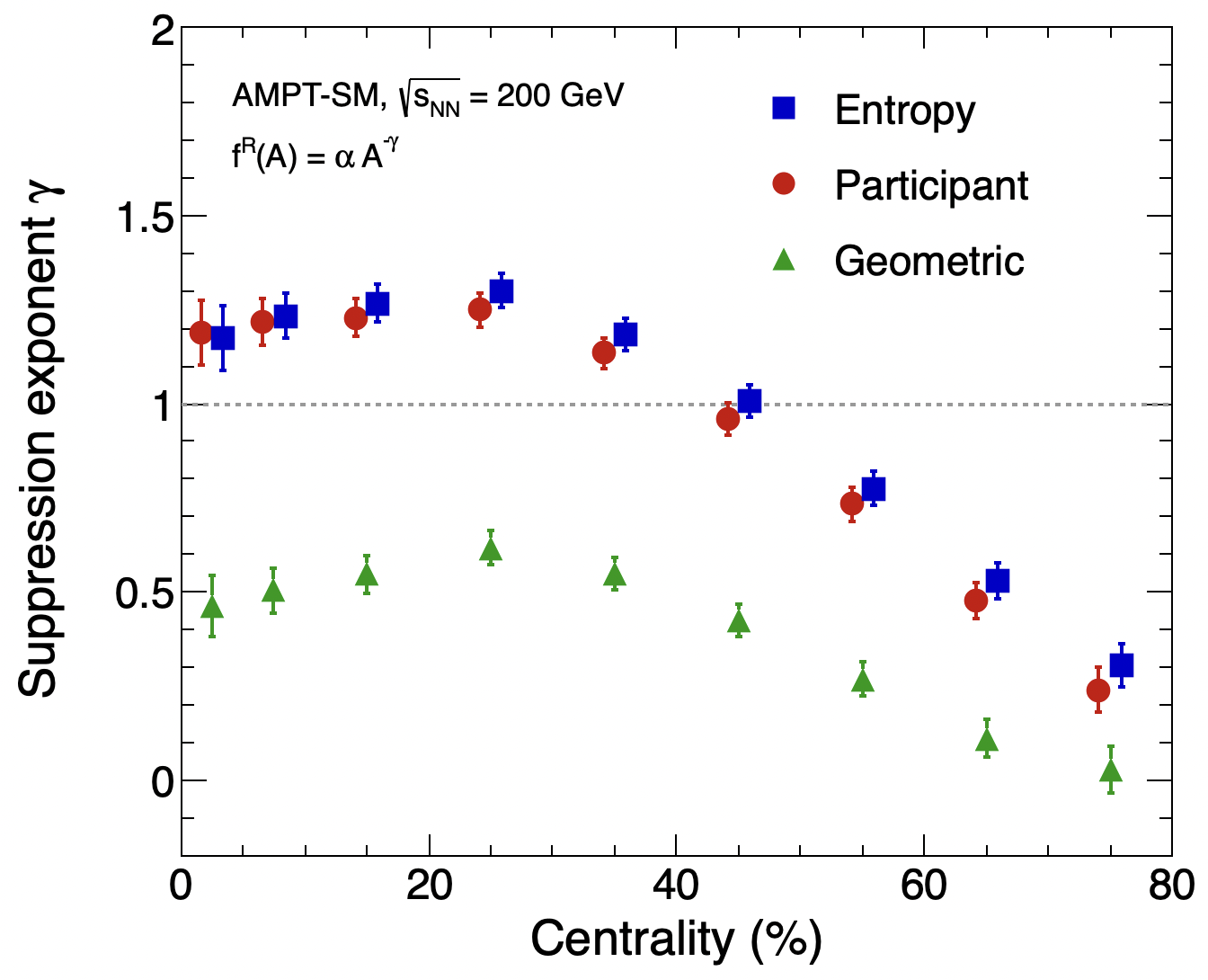}
\caption{The suppression exponent $\gamma$ as a function of collision centrality (\%), obtained from the fit to the observable $R(A)$ with the Power-law function of Eq.~\ref{eq:powerlawF}, at $\snn =$ 200 GeV from the AMPT-SM model. The markers in each centrality bin are slightly shifted from the central position for visual clarity.}
\label{fig:gamma-cent}
\end{figure}

Figure~\ref{fig:gamma-cent} shows the suppression exponent $\gamma$ for the entropy, participant, and geometric scaling observables. These values are extracted using the fit to the three scaling observables with power-law function presented in Eq.~\ref{eq:powerlawF} for narrow centrality intervals from 0-5\% to 70-80\%. The values of $\gamma$ for entropy and participant scaling are consistent within uncertainties across all the centrality classes. The extracted $\gamma$ shows an increase from the most peripheral to mid-central collisions, reaching a maximum at 20-30\% centrality interval and then saturates for central collisions (0-20\%). 
The $\gamma$ remains approximately constant ($\gamma\approx$ 1.2) for the entropy and participant scaling in the central region, and crosses $\gamma =1$ limit at centrality 40-50\%. A value of $\gamma$ greater than unity indicates a stronger suppression of the per-source flow signal as the system size increases. The exponent $\gamma$ in the case of geometric scaling is quite different than entropy and participant scaling. Its value is much smaller, suggesting that simple geometric scaling of the studied flow observable does not fully account for the system-size dependence of the medium. 

It is important to note that the suppression exponent $\gamma$ is obtained using a simple power-law functional form of the scaling observables $S(A)$, $P(A)$, and $G(A)$, as shown in Fig.~\ref{fig:RSD3x3}. Since these three observables differ in magnitude, it is necessary to normalize them to draw any quantitative conclusions. The above approach is thus our primary motivation for constructing a dimensionless ratio $\xi(A)$ to quantify the threshold ($\AAcrit$) at which the collectivity of the medium changes.  

\subsection{The Ratio $\xi(A)$ and Extraction of $\AAcrit$}
\label{sec:Acrit}
\begin{figure*}[!htbp]
\centering
\includegraphics[width=0.95\textwidth]{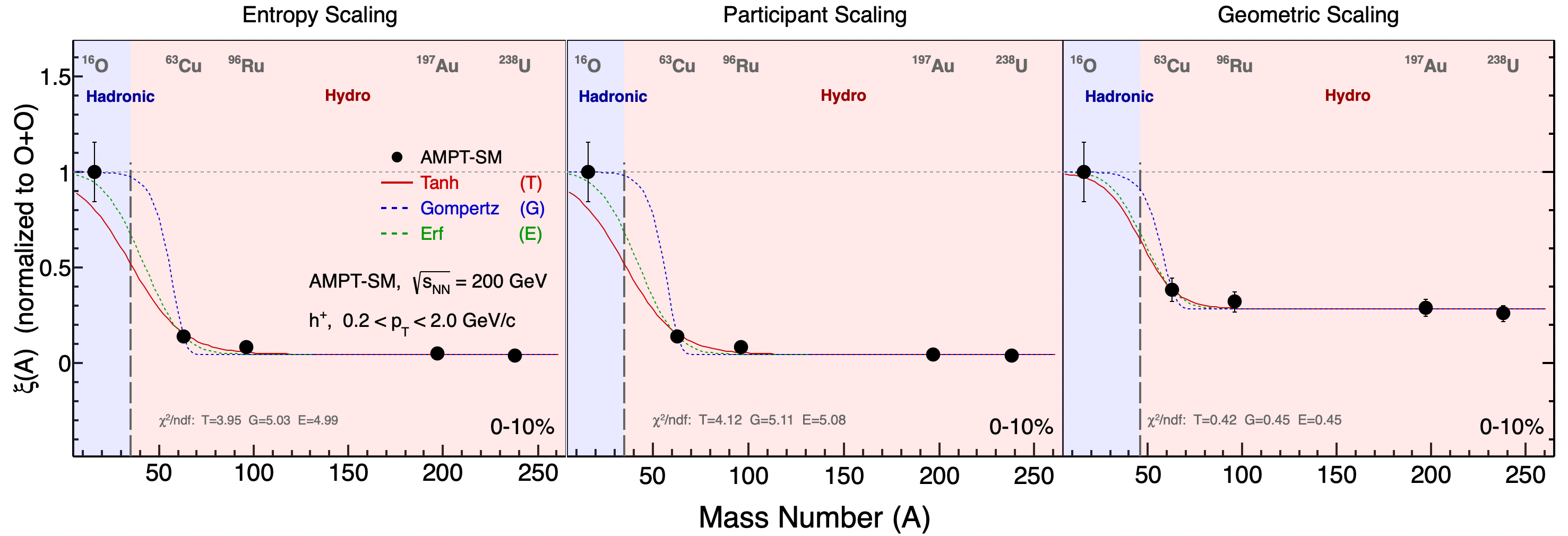}
\caption{The ratio $\xi(A)$ for the three scaling observables in central collisions (0-10\%) at $\snn =$ 200 GeV from the AMPT-SM model. The solid lines represent the fits using the three S-curves: Tanh, Gompertz, and Error function. The threshold mass numbers associated with these three S-curves are denoted by $\AAcrit^{T}$, $\AAcrit^{G}$, and $\AAcrit^{E}$. The $\chi^2/ndf$ values for the three fits are also mentioned. The blue and red shaded areas mark the hadronic and hydrodynamic-dominated regimes, separated by $\AAcrit^{T}$.}
\label{fig:xi-fits}
\end{figure*}
Figure~\ref{fig:xi-fits} shows the ratio $\xi(A)$ for the entropy, participant, and geometric scaling observables in central (0-10\%) collisions at $\snn =$ 200 GeV from the AMPT-SM model. The fitting curves represent Tanh, Gompertz, and Error function discussed in Sec.~\ref {sec:framework-fits}. The mass number at which the function reaches half-maximum is referred to as the threshold mass number, $\AAcrit$. In the region, above $\AAcrit$, $\xi(A)$ remains constant while it exhibits an abrupt increase below $\AAcrit$. According to Eq.~\ref{eq:RA}, the region below $\AAcrit$ is then identified as ``hadronic'', while the region above is identified as ``hydrodynamic''. For central collisions, the $\AAcrit$ values from the Tanh, Gompertz, and Error functions agree within their uncertainties. The two regions separated by $\AAcrit$ show similar behavior in the case of entropy and participant scaling, whereas geometric scaling exhibits higher $\AAcrit$ values.
\begin{figure*}[!htbp]
\centering
\includegraphics[width=0.9\textwidth]{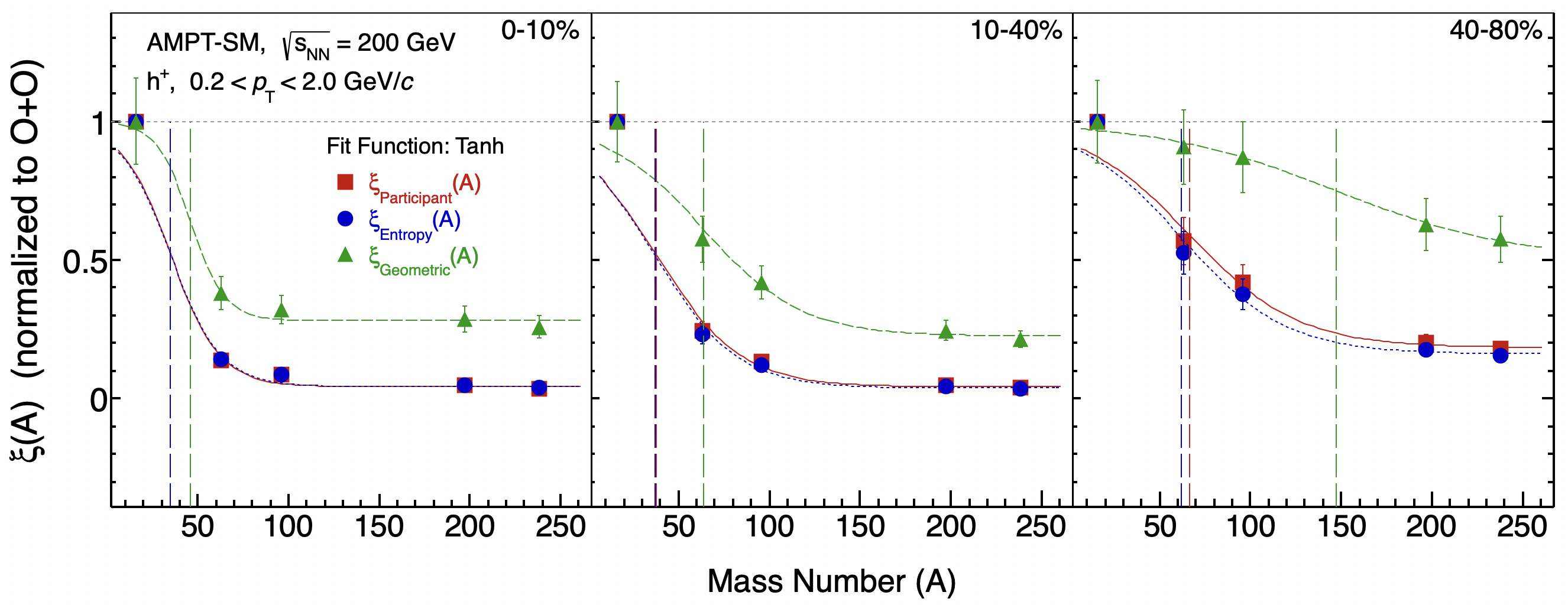}
\caption{The ratio $\xi(A)$ in 0-10\%, 10-40\%, and 40-80\% centrality classes from the AMPT-SM model at $\snn =$ 200 GeV. The data points are fitted using the Tanh function. The extracted values of $\AAcrit$ are represented by vertical dashed lines.}
\label{fig:xi-overlay}
\end{figure*}

Figure~\ref{fig:xi-overlay} shows the ratio $\xi(A)$ for the three scaling observables, calculated using Eq.~\ref{eq:xiSC} in different centrality intervals 0-10\%, 10-40\%, and 40-80\% from the AMPT-SM model at $\snn =$ 200 GeV. The three observables, $\xiE$, $\xiP$, and $\xiG$ are fitted with the Tanh function as described in Eq.~\ref{eq:tanhfit}. The values of threshold mass number, $\AAcrit$, in each case are extracted from the fit function at the point where $\xi(A)$ crosses the half-maximum. The extracted $\AAcrit$ values for the three scaling approaches and different centrality intervals are summarized in Table~\ref{tab:Acrit-summary}.

The $\xiE$ and $\xiP$ curves are statistically indistinguishable from each other, despite the different underlying denominators, $\dSdy$ and $\Npart$. In contrast, the $\xiG$ curve shows a different mass number dependence. The $\AAcrit$ values for entropy and participant scaling are comparable within uncertainties across all centrality intervals. However, $\AAcrit$ is systematically higher for geometric scaling compared to the other scalings. This indicates $\dSdy$ and $\Npart$ characterize the amount of matter produced in the collisions, whereas $A^{1/3}$ characterizes its transverse extend, so the two families of scaling variables set the threshold by different measures of the system size. The threshold mass number, $\AAcrit$, places the onset of collective behavior between mass number $A=16$ and $A=63$. The experimental data available in this mass number range can be used to test the onset of collectivity and the applicability of a hydrodynamic description~\cite{hydro_quantify, hydro_review}. 

\begin{table}[!htbp]
\caption{Threshold mass number $\AAcrit$ from the Tanh fit for the three scaling observables and centrality intervals. The errors represent statistical uncertainties from the fit.}
\label{tab:Acrit-summary}
\begin{ruledtabular}
\begin{tabular}{lccc}
Centrality    & Entropy        & Participant    & Geometric \\
\hline
$0$--$10$\%   & $34.8 \pm 7.7$ & $34.9 \pm 7.8$ & $46.1 \pm 16.8$\\
$10$--$40$\%  & $36.8 \pm 6.5$ & $37.6 \pm 6.6$ & $63.6 \pm 9.0$\\
$40$--$80$\%  & $61.9 \pm 8.3$ & $66.3 \pm 8.9$ & $147.3 \pm 32.3$\\
\end{tabular}
\end{ruledtabular}
\end{table}

In peripheral collisions, the values of $\AAcrit$ are higher compared to those in the most central collisions. Further, the $\AAcrit$ value obtained for $\xiG$ in peripheral collisions becomes exceptionally high ($\AAcrit = 147.3 \pm 32.3$). Such a high threshold mass number again indicates that geometric scaling is broken. In other words, a simple geometric scaling is insufficient, and the normalization fails to quantify the region where the transition from hadronic to hydrodynamic is expected to occurs.
\begin{figure}[!htbp]
\centering
\includegraphics[width=0.96\columnwidth]{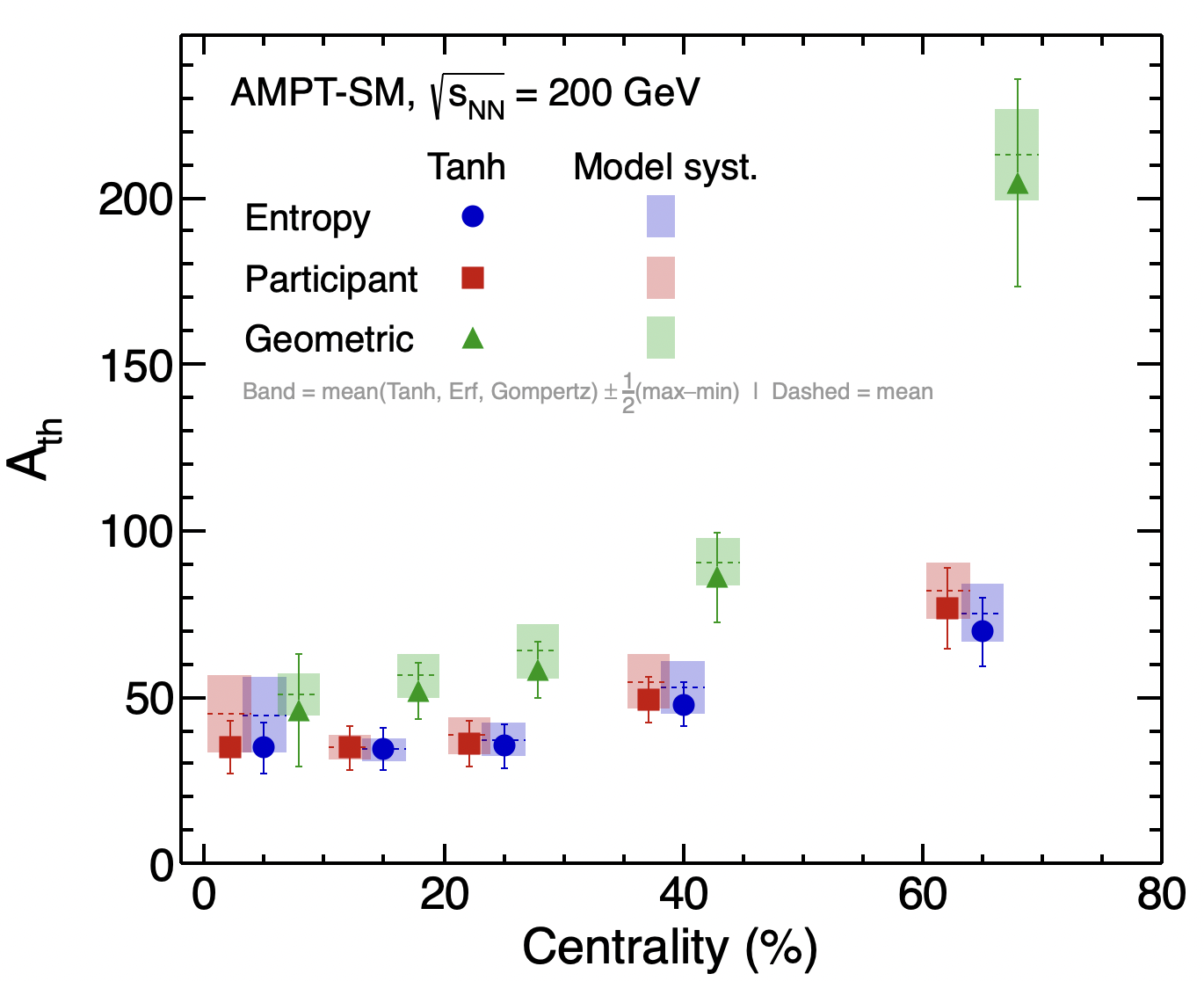}
\caption{Threshold mass number ($\AAcrit$) as a function of collision centrality (\%) for the three scaling observables at $\snn =$ 200 GeV from the AMPT-SM model. The colored bands represent systematic uncertainties from the model. Horizontal dashed line at each centrality shows the mean of $\AAcrit$ from the three s-curves fitting. Symbols are slightly offset along x-axis for better visual clarity.}
\label{fig:Acrit-cent}
\end{figure}

Figure~\ref{fig:Acrit-cent} summarizes the results of threshold mass number as a function of collision centrality for the three scaling approaches at $\snn =$ 200 GeV from the AMPT-SM model. The default value of $\AAcrit$ is obtained from the Tanh function and is represented as solid markers. The mean values of $\AAcrit$, represented by horizontal dotted line, are calculated by averaging the results from the Tanh, Gompertz, and Error functions. The systematic uncertainties on the mean values are determined as the difference between the maximum and minimum values of $\AAcrit$ from the three fitting functions. In the central collisions (below 30\%), the $\AAcrit$ remains constant at a value of about 35 for both entropy and participant scaling. This central plateau is the most significant and robust feature of the present analysis, suggesting that in the hydrodynamic regime, $\AAcrit$ is determined by the system's thermodynamic threshold rather than its centrality. In peripheral collisions (30-80\%), both $\AAcrit^{\,\mathrm{Entropy}}$ and $\AAcrit^{\,\mathrm{Participant}}$ gradually increase, reaching about 70. The $\AAcrit^{\,\mathrm{Geometric}}$ follows a similar qualitative trend but has higher values, reaching about 200 in the most peripheral collisions, showing that geometric scaling is effectively broken. 
 
\subsection{Knudsen-number map and the driver-damping of the directed-flow}
\label{sec:Kn-v1}
\begin{figure}[!htbp]
\centering
\includegraphics[width=0.96\columnwidth]{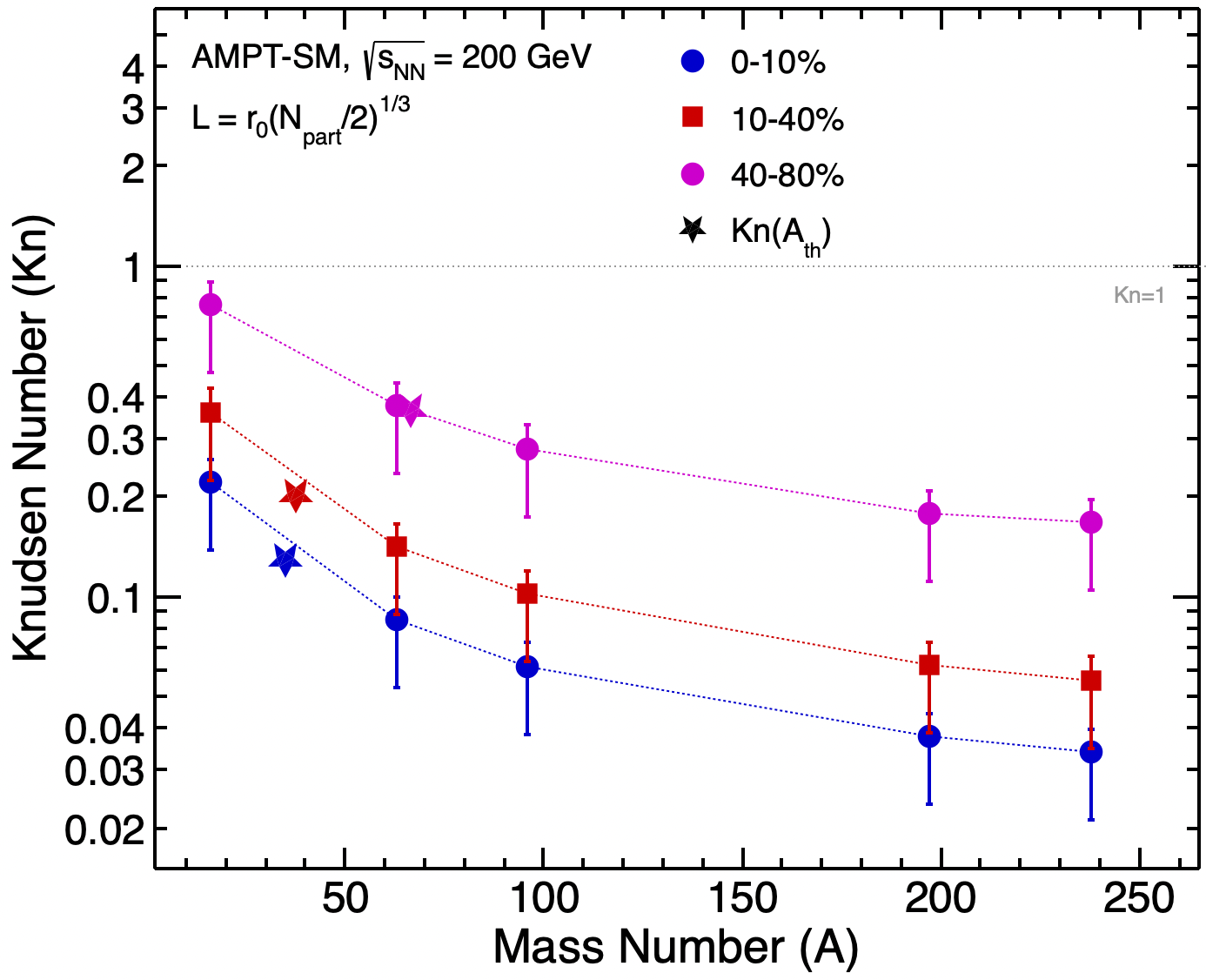}
\caption{Kinetic-theory Knudsen number $\Kn$  as a function of $A$ for the three centrality classes.  The horizontal dotted line at $\Kn = 1$ marks the nominal boundary between the hydrodynamic ($\Kn < 1$) and the transport ($\Kn > 1$) regimes.  Star markers indicate the value of $\Kn$ at the data-driven $\AAcrit$ for each centrality bin.}
\label{fig:Kn}
\end{figure}

Figure~\ref{fig:Kn} shows the Knudsen number ($\Kn$) computed using Eq.~\ref{eq:Kn} as a function of mass number ($A$) for three centrality intervals: 0-10\%, 10-40\%, and 40-80\% at $\snn =$ 200 GeV from the AMPT-SM model~\cite{Bjorken:1982qr,Drescher:2007cd}. The calculation uses the parton-parton cross-section of $\sigma_p = 1.5$ mb~\cite{He:2017tla}. The charged particle multiplicity $\dNdeta$ within the pseudo-rapidity range of $|\eta| < 0.5$ is used. The effective fireball radius $L$ is obtained using the $\Npart$ information from the AMPT-SM model for each of the collision systems. The star markers in Fig.~\ref{fig:Kn} indicate the $\Kn$ values at the threshold mass number for each centrality interval. The $\Kn(\AAcrit)$ values are about $0.08$, $0.15$, and $0.27$ for central, mid-central, and peripheral collisions, respectively. Systematic uncertainties on $\mathrm{Kn}$ are evaluated by varying the parton multiplicity per hadron over $f_p \in [2,3]$ and the formation time over $\tau_0 =$ 0.5 to 0.7~fm/$c$. 

The values of $\Kn$ for heavier collision systems (U+U and Au+Au) are significantly below unity. As the system size decreases, these values increase and approach the reference value of $\Kn = 1$. This indicates that the medium produced in smaller collision systems behaves like a non-interacting hadronic medium. The $\Kn$ values also show a clear dependence on centrality, increasing from most-central to peripheral collisions. This is expected, as central collisions have higher multiplicity and more number of participants. Notably, the $\Kn$ values ($\approx$ 0.2) in central O+O collisions are comparable to the peripheral Au+Au or U+U collisions. This suggests that the microscopic mean free path decreases more rapidly than the effective macroscopic length.  

The collision systems studied in this work at $\snn = 200$~GeV using the AMPT-SM model lie within the hydrodynamic regime, as indicated by Knudsen number below 1 ($\Kn <$ 1), according to the kinetic theory approach. The value of $\Kn(\AAcrit)$ increases by a factor of three from central to peripheral collisions, showing a centrality dependence similar to that reflects in $\AAcrit$ as shown in Fig.~\ref{fig:Acrit-cent}. Therefore, the increase in $\Kn(\AAcrit)$ towards the non-hydrodynamic regime suggests a decrease in the collectivity of the medium produced in relativistic nucleus-nucleus collisions at $\snn = $ 200 GeV, within the framework of the AMPT-SM model. The quantity $\AAcrit$ is a data-driven observable and can be obtain using the directed flow slope from the experimental data. 

\begin{figure}[!htbp]
\centering
\includegraphics[width=0.96\columnwidth]{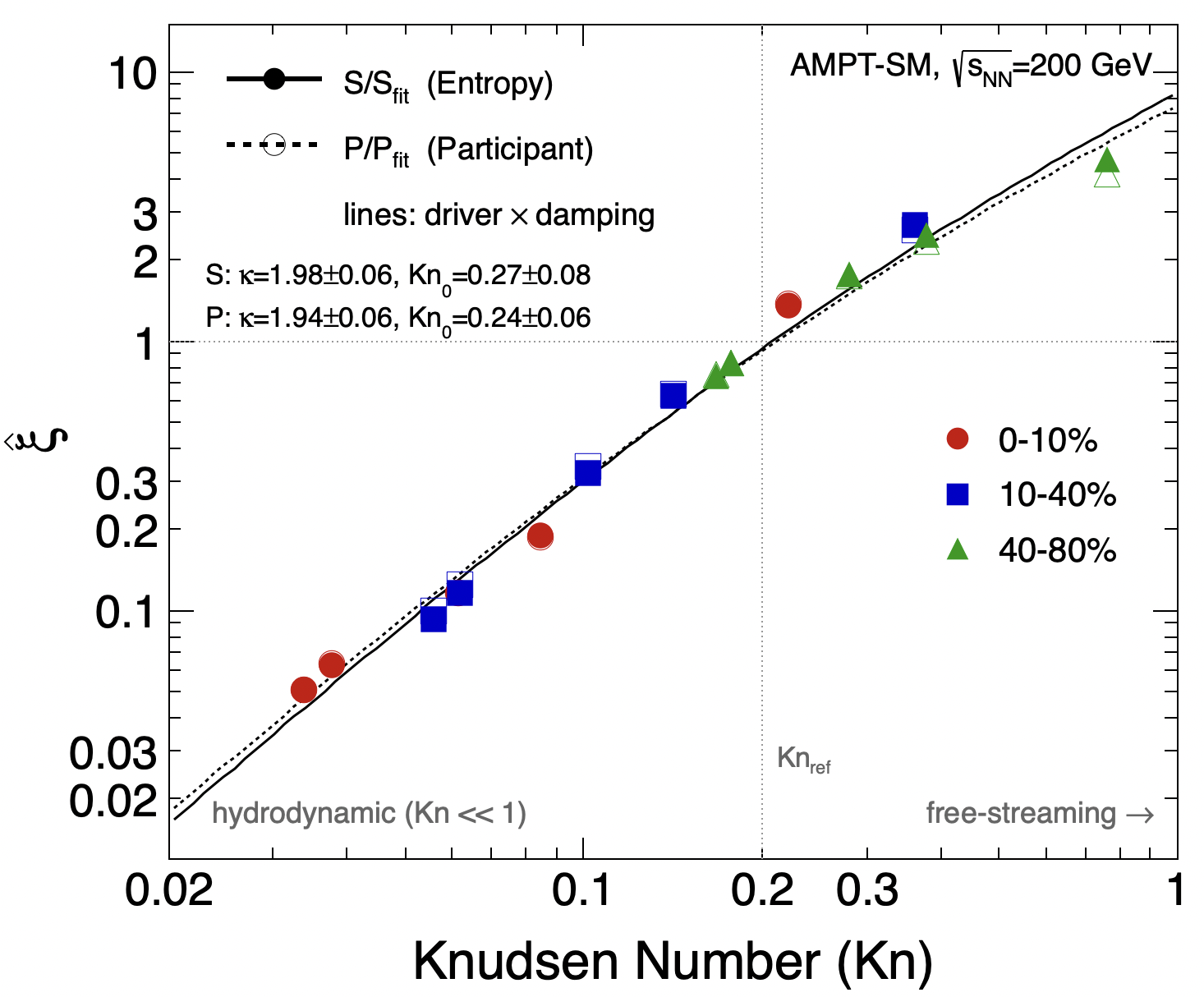}
\caption{Normalized $v_1$ response $\hat{\xi} = F/F_{\mathrm{fit}}(\Kn_{\mathrm{ref}})$ versus Knudsen number $\Kn$, for participant-scaled (solid markers) and the entropy-scaled (open markers) $v_1$-slope in different centrality intervals and collision systems. The two curves (solid and dotted lines) represent the global fit to the $\hat{\xi}$ with Eq.~\ref{eq:driverdamping}.}
\label{fig:xi-Kn}
\end{figure}
The suppression exponent $\gamma$ as shown in Fig.~\ref{fig:gamma-cent} suggests two distinct sources that contribute to the system-size dependence of the medium produced in relativistic nucleus-nucleus collisions. The first source is the initial-state \emph{driver} of $v_1$, stemming from the tilted, dipole-like geometry of the fireball~\cite{Adil:2005qn, Bozek:2010aj, Teaney:2010vd}. The second source is final-state viscous \emph{damping}, which affects the hydrodynamic response of the medium. Viscous corrections to the ideal hydrodynamics enter through observables $\Kn \propto (\eta/s)/L$, where $\eta/s$ represents the ratio of shear viscosity to entropy density, and $L$ is a characteristic macroscopic length scale. By studying the scaled slope as a function $\Kn$ instead of the mass number $A$, we can separate the contributions from the two sources, initial-state \emph{driver} and final-state \emph{damping}.

Figure~\ref{fig:xi-Kn} shows the entropy-scaled $v_1$-slope $S$, normalized within each centrality group to its power-law interpolation at a common reference point, $\hat{\xi} = S/S_{\mathrm{fit}}(\Kn_{\mathrm{ref}})$ with $\Kn_{\mathrm{ref}} = 0.2$. Here, $S_\mathrm{fit}$ is the fit to $S$ using the power-law function $f^{dr} \propto \Kn^{\kappa}$. The participant-scaled $v_1$-slope $P$ is also presented to show independence of normalization in contrast to the ratio $\xi(A)$ in Eq.~\ref{eq:xiSC}, which uses an extensive normalization based on the O+O collisions. The comparison shown in Fig.~\ref{fig:xi-Kn} across different collision centralities and systems requires a common reference point $\Kn_{\mathrm{ref}}$. The choice of $\Kn_{\mathrm{ref}}$ is a pure convention, and leads to the $\hat{\xi}(\Kn_{\mathrm{ref}})=1$.

The central group follows a power-law dependence with statistically similar exponents $\kappa = 1.73 \pm 0.06$ and $1.74 \pm 0.03$ for the $0-10\%$ and $10-40\%$ centralities, respectively. The peripheral group breaks the universality with a much shallower exponent, $\kappa = 1.14 \pm 0.04$. A damping-only dependence, described by the equation $f^{\mathrm{dm}} \propto 1/(1 + \Kn/\Kn_0)$~\cite{Bhalerao:2005mm,Drescher:2007cd,Gombeaud:2007ub}, fails to accurately replicate either the sign or the magnitude of the observed trend in all the centrality groups. The system-size dependence of the scaled $v_1$-slope is therefore dominated by its initial-state driver component rather than by viscous attenuation. 

To achieve a consistent global fit across all three centrality groups and systems, we employ a function that includes both the driver and damping components. The function is defined as:
\begin{equation}
  f^{dr+dm} \;=\; C\,  \frac{(\Kn/\Kn_{\mathrm{ref}})^{\kappa}}{1 + \Kn/\Kn_0}.
  \label{eq:driverdamping}
\end{equation}
A fit to entropy-scaled slope $S$ yields, \[\kappa = 1.98 \pm 0.06, \qquad \Kn_0 = 0.27 \pm 0.08,\] with the advantage of reduction in $\chi^2/ndf$ from 13 (using only power-law component) to 7 (including both components), at the expense of one additional parameter. Repeating the global fit to the participant-scaled slope $P$ yields statistically similar
parameters, \[\kappa = 1.94 \pm 0.06, \qquad \Kn_0 = 0.24 \pm 0.06,\] with the $\chi^2/ndf$ reduced from 15 to 8. The uncertainties on the extracted parameters represent uncertainties from the fit to $\hat{\xi}$ using Eq.~\ref{eq:driverdamping}.

The two parameters carry distinct physics. The driver component characterizes the initial push of the medium caused by a tilted dipole-like geometry. The exponent $\kappa \simeq 2$ implies that the initial-state driver of the scaled $v_1$-slope increases approximately quadratically with $\Kn$. This indicates a stronger initial tilt per unit entropy toward the smaller systems. 

The damping component reflects the conversion of the initial-state push into the directed flow of particles in the final-state momentum space. A fluid transmits this push efficiently when its constituents collide sufficiently; in other words, when the mean free path is small relative to the size of the system. In such conditions, particles in the medium interact and build up the flow. The damping scale $\Kn_0 =$ 0.27 indicates the value of the Knudsen number ($\Kn$) at which the conversion of the initial state geometry into directed flow retains only half of its ideal hydrodynamic efficiency.
At $\Kn_0 =$ 0.27, the fraction $(1+\Kn/\Kn_0)^{-1}$ is between 0.55 and 0.26 for values of $\Kn$ ranging around 0.22 to 0.76. As a result, in O+O collisions, the response of the medium is suppressed by 45\% (0-10\%) to 75\% (40-80\%). In comparison to small O+O collisions, the central Au+Au and U+U collisions, where $\Kn$ is approximately 0.03 to 0.04, the response is retained about 80\% to 90\% of the ideal hydrodynamic response. The damping scale of directed flow $\Kn_0 \approx 0.27$ is about a factor of 2.5 smaller than the elliptic flow ($v_2$) value $\Kn_0 \approx 0.7$ extracted from the $v_2$ measurements from the experimental data at the RHIC~\cite{Drescher:2007cd,Gombeaud:2007ub}. Therefore, the hydrodynamic response of the medium reduces $v_1$ earlier than that of $v_2$. This suggests that directed flow is more sensitive to incomplete equilibration than the elliptic flow.

The damping scale $\Kn_0$ can be used to estimate the shear viscosity to entropy density ratio as an independent cross-check of its kinetic-theory estimator. By inverting the equation, $\Kn = [5\hbar c\,(\eta/s)]/(T L)$, and evaluating it at $\Kn_0$ gives $(\eta/s)_{\mathrm{eff}} = (\Kn_0 TL)/(5\hbar c) \approx 0.15 \pm 0.05$ for O+O collisions with the conditions $T \approx 0.22 \pm 0.03$~GeV and $L \approx 2.4 \pm 0.4$~fm~\cite{Kolb:2003dz,Heinz:2013th,PRL.104.132301}.

\section{Summary and outlook}
\label{sec:summary}
We studied system-size dependence of charged-hadron directed flow slope in $^{16}$O+$^{16}$O, $^{63}$Cu+$^{63}$Cu, $^{96}$Ru+$^{96}$Ru, $^{197}$Au+$^{197}$Au, and $^{238}$U+$^{238}$U collisions at $\snn = 200$~GeV, using an improved AMPT-SM model~\cite{He:2017tla}. We defined three scaling observables for the directed flow slope based on the entropy density, the number of participants, and the mass number. Further, we constructed a dimensionless ratio $\xi(A)$ to extract the threshold mass number $\AAcrit$ by fitting it with a family of three S-curves (Tanh, Error, and Gompertz). The threshold mass number is found to be $A \approx 35$ for entropy and participant scaling in central collisions (0-40\%). The corresponding $\AAcrit$ from geometric scaling is much higher, indicating that simple mass number scaling of the directed flow slope does not hold. 

A Knudsen number map is obtained based on the kinetic theory approach from the AMPT-SM model at a given collision system and centrality interval. The driving and the damping contributions are separated by plotting the scaled directed flow slope against $\Kn$. The results show that the initial-state driver component of the scaled directed flow increases towards the lighter systems. The damping-only component cannot account for the observed sign or magnitude of the scaled directed flow slope. Therefore, the increase in the directed-flow slope is attributed to the initial-state driver, which arises from the initially tilted dipole-like geometry of the colliding systems. A global fit that incorporates both the driver and damping components gives $\kappa = 1.98 \pm 0.06$ and $\Kn_{0} = 0.27 \pm 0.08$ for entropy scaling, and $\kappa = 1.94 \pm 0.06$ and $\Kn_{0} = 0.24 \pm 0.06$ for participant scaling. The damping scale $\Kn_{0}$ represents the Knudsen number at which half of the ideal hydrodynamic response is preserved. The value of $\Kn_{0} \approx 0.27$ is about a factor of 2.5 smaller than $\Kn_{0} \approx 0.7$ reported from the elliptic flow~\cite{Drescher:2007cd}. Furthermore, an effective shear viscosity to entropy density ratio, $(\eta/s)_{\mathrm{eff}} \approx 0.15 \pm 0.05$, is obtained from the kinetic-theory Knudsen-number map without requiring fitting to the magnitude of the flow harmonics.

A follow-up to the current analysis can be performed to achieve a more precise identification of the onset of collectivity based on the identified hadrons. The threshold mass number, which is extracted separately for pions, kaons, and protons, may differ because these hadron species receive different contributions from radial flow, hadronic rescattering, and baryon transport, respectively. By comparing the values of $\AAcrit^{\pi}$, $\AAcrit^{K}$, and $\AAcrit^{p}$, we can determine whether the threshold is a general characteristic of the bulk medium or if it depends on the hadron species.

The same procedure can be applied to the higher-order flow harmonics. Elliptic and triangular flow are driven by the eccentricity and the triangularity of the initial state, and their damping scales may not be similar to that of directed flow. Extracting $\AAcrit$ and $\Kn_{0}$ for $v_{2}$ and $v_{3}$ using the same scaling observables would establish that a single threshold describes the onset of collective behavior. Such a comparison would also test the driver-damping decomposition in a harmonic-dependent way, since the driver exponent $\kappa$ reflects the initial-state geometry, which differs from one harmonic to another, while $\Kn_{0}$ reflects the dissipative response of the medium. The data-driven approach outlined in this work can be applied directly to experimental data on various collision systems at the RHIC and LHC~\cite{ALICE:OO_NeNe,ATLAS:2026OONeNe}. It allows the determination of a threshold mass number at which collectivity begins. Measurements of these quantities in the experimental data help constrain the threshold mass number and provide a connection between the hydrodynamic and transport-dominated regimes of matter produced in relativistic nucleus-nucleus collisions.

\begin{acknowledgments}
KN is supported by OSHEC, Department of Higher Education, Government of Odisha, Index No.\ 23EM/PH/124 under MRIP 2023.  The authors thank Prof.\ B.~Mohanty for providing computational facilities at
NISER, India, and Prof.\ Z.-W.\ Lin for the new coalescence AMPT model.
\end{acknowledgments}

\bibliographystyle{apsrev4-2}
\bibliography{bibliography}

\end{document}